%Paper: hep-th/9303137
%From: chou@physics.rockefeller.edu
%Date: Wed, 24 Mar 1993 14:39:25 EST

%%%%%%%%%%%%%%%%%%% text of paper %%%%%%%%%%%%%%%%%%%%%%%%%%%%%%%%%%%%%%%%%%%%
\def\nolabels{\def\eqnlabel##1{}\def\eqlabel##1{}\def\reflabel##1{}}
\def\writelabels{\def\eqnlabel##1{%
{\escapechar=` \hfill\rlap{\hskip.09in\string##1}}}%
\def\eqlabel##1{{\escapechar=` \rlap{\hskip.09in\string##1}}}%
\def\reflabel##1{\noexpand\llap{\string\string\string##1\hskip.31in}}}
\nolabels
\global\newcount\meqno \global\meqno=1
\global\meqno=1
\def\eqnn#1{\xdef #1{(\the\meqno)}%
\global\advance\meqno by1\eqnlabel#1}
\def\eqna#1{\xdef #1##1{\hbox{$(\the\meqno##1)$}}%
\global\advance\meqno by1\eqnlabel{#1$\{\}$}}
\def\eqn#1#2{\xdef #1{(\the\meqno)}\global\advance\meqno by1%
$$#2\eqno#1\eqlabel#1$$}
\overfullrule=0pt
\magnification=\magstep1
\font\twelvebf=cmbx12
\nopagenumbers
\footline={\ifnum\pageno>1\hfil\folio\hfil\else\hfil\fi}
\pageno=1
\line{\hfil RU-93-20-B}
\line{\hfil March 1993}
\vglue .7in
\centerline{{\twelvebf On the Quantizations of the Damped
Systems}\footnote{*}
{This work is
supported in part by funds provided by the U.S. Department of Energy (D.O.E.)
under contract \#DOE91ER40651.B}}
\vskip .4in
\centerline{\it Chihong Chou}
\vskip .1in
\centerline { Physics Department, Rockefeller University}
\centerline {New York, New York 10021, U.S.A.}
\vskip .2in
\vskip .5in
\baselineskip=14pt
\centerline {\bf Abstract}
\vskip .2in
Based on a simple observation that a classical second order differential
equation may be decomposed into a set of two first order equations, we
introduce a Hamiltonian framework to quantize the damped
systems. In particular, we analyze the system of a linear damped
harmonic oscillator and demonstrate that the time evolution of the
Schr$\ddot{\rm o}$dinger equation is unambiguously determined.
\vskip .6in
\noindent
%This work was supported in part by the US Department of Energy.
\vfill
\eject
\pageno=2
\baselineskip=24pt
The search for field theoretic formulations of
fundamental anyons$^{1,2}$ has refocused attention on
the long outstanding problem of the quantization
of (non-Hamiltonian) systems$^{1-7}$ in which only the equation of motion
is explicitly known. The common feature of this class of systems
is its history dependence.
The most notable such system is certainly that of
the damped harmonic oscillator. Apart from requiring the knowledge
of quantum aspects of this class of systems for practical reasons
(for example, the laser system), this problem has its
own right to be investigated both for the mathematical interests
and exploring the new (perhaps deeper) fundamental features
of the quantum theory.

There have been quite a number of attempts to solve this problem, which
may be classified into two classes. One is to start with
a modified quantization scheme$^{4,5}$ or a bigger system with introduction
of additional degrees of freedom$^{6}$ such that the conventional
Schr${\ddot{\rm o}}$dinger
or Heisenberg dynamical description is valid. The other$^{7}$
is to formulate the quantum theory in a modified Schr${\ddot{\rm o}}$dinger
dynamics (a nonlinear one). Clearly, attempts in the second
class do not respect the superposition principle as the
nonlinearity enters as a consequence of wavefunction dependent
potentials. Similarly, the first class of the attempts also has
some disadvantages. The introduction of the non-hermitian or
time-dependent Hamiltonian in this class often leads to the
conclusion of non-existence of the Schr$\ddot{\rm o}$dinger wave
description.$^{4}$

As is usually done in quantum mechanics, the time evolution
of the Schr$\ddot{\rm o}$dinger dynamics may be carried out only if
one insures the completeness of the Hamiltonian eigenstates. The
hermitian Hamiltonian in our standard theory guarantees
unique unitary time evolution. For dissipative systems, the
time evolution is certainly no longer unitary, yet one may
hope that the Schr$\ddot{\rm o}$dinger quantum mechanics
may be still used for unambiguously determining the time evolution.
The purpose of this paper is to present an alternative
way of modifying the quantization framework in which
the Lagrangians are introduced naturally and to show that
the Schr$\ddot{\rm o}$dinger
dynamics is indeed achieved. Thus the above mentioned puzzle
will be resolved.\footnote{$^\#$}{During the preparation
of this paper, we receive a paper by Feshbach and Tikochinsky,$^{5}$ in which
this puzzle was also resolved in a different way. We thank Roman Jackiw
for bringing this reference to our attention.}

We shall begin with a general damped system on the line
in an external potential,
whose classical motion is govern by the non-linear second order
differential equation:
\eqn\a{\ddot{x} +k(x)\dot{x}+g(x)=0,}
where the dots denote time derivatives and $k(x)$ and $g(x)$
are real functions. We observe that a reduction of Eq. \a \ to
\eqn\b{\dot{x}+f(x)=0}
may be defined if $f$ satisfies
\eqn\c{f(f'-k)+g=0,}
where the prime represents an $x$-derivative. Note the complex
conjugate $f^*$ of $f$ will be also a solution of Eq. \c \ if $f$
is a solution.
One may verify this reduction by directly differentiating Eq. \b \
with respect to time and using of the condition \c.
Roughly speaking, this reduction reflects
the fact that a second order differential
equations may be decomposed into a set of two first order equations.
For the obvious reasons, this reduction has at least two advantages.
Classically, a first order equation (linear or not) is of course simpler than
the second order one,\footnote{*}{These ideas may be used to solve the
sine-Gordon and other nonlinear equations.$^{8}$} while quantum mechanically,
the first order equations may be regarded as the Hamiltonian
equations by properly defining the canonical variables.
For later convenience, we denote $x_i(i=1,2)$ as the two (independent)
solutions of Eq. \a \ or the solutions of the follow set of the first
order equations:
\eqn\d{\dot{x_i}+f_i(x_i)=0,}
where $f_i$ are two (independent) solutions of Eq. \c. One may easily
show that $x=x_1+x_2$ will be a solution of Eq. \a \ only for the linear
damped case, namely, $k$=constant and $g$ is a linear function of $x$.

The Lagrangians $L$
\eqn\x{L=\sum_{i=1}^2 [\dot{x_i}p_i-x_i\dot{p_i}+2f_ip_i],}
which may be naturally introduced and give rise to Eq. \c \ and
\eqn\f{\dot{p_i}-{1\over 2}(f_i'(x_i)p_i+p_if_i'(x_i))=0,}
where no summation is assumed for the repeat
index and the symmetric form
is used for the usual reason. Thus the Hamiltonian
is obtained in the usual manner (we use the symmetric form to
deal with the ambiguous quantum ordering):
\eqn\g{H=-\sum_i (f_ip_i+p_if_i).}
The symplectic two-form for the system is
\eqn\h{\omega=2\sum_i dp_i\wedge dx_i,}
which defines the Poisson brackets. Therefore, the commutation
relations are
\eqn\y{\eqalign{&[x_i, x_j]=[p_i, p_j]=0 \cr
		&[x_i, p_j]={i\over 2}\hbar \delta_{ij}.}}
Using the rules \y \ and the Hamiltonian \g, one may easily verify
that the classical equations \d \ and \f \ are quantum mechanically
realized (up to the quantum corrections $O(\hbar)$)
as the Heisenberg equations:
\eqn\j{\eqalign{&\dot{x_j}={1\over i\hbar}[x_j, H], \cr
		&\dot{p_j}={1\over i\hbar}[p_j, H].}}
Note that $x_i$ and $p_i$ may not be considered as hermitian operators
and in the case of $f_1=f_2^*$ the consistency condition requires
the identifications $x_1=x_2^*$ and $p_1=p_2^*$ classically(in quantum case,
they are $x_1=x_2^\dagger$ and $p_1=p_2^\dagger$). We will keep this
in mind for later discussion.

Now let us analyze the case of a linear damped harmonic oscillator,
namely $k=2\lambda\ge 0$ and $g(x)=\omega^2 x$(mass=1). Thus we have
$f=-\eta x$ from Eq. \c, where $\eta_i=-\lambda\pm \sqrt{\lambda^2-
\omega^2}$[$\eta_{1(2)}$ with the upper (lower) sign]
with the convention $\eta_1\ge\eta_2$ (or Im $\eta_1\ge$ Im
$\eta_2$), if $\eta_i$ are real (or complex). Namely, the square root
in this paper will be always positive and $\sqrt{\lambda^2-
\omega^2}=i\sqrt{\omega^2-\lambda^2}$ for $\lambda<\omega$.
The equations of motion (4) and (6) then become
\eqn\l{\eqalign{&\dot{x_i}-\eta_i x_i=0, \cr
		&\dot{p_i}+\eta_i p_i=0.}}
The $p$-system is clearly related to the $x$-system
by a time-reversal transformation
$U_t:\ t\rightarrow -t$, or equivalently $U_{\lambda}:\
\lambda\rightarrow -\lambda$.
Note that since the complex index $i$ or the
square root is introduced
in our reduction, the time reversal transformation
has been identified as the complex conjugation or simply
shifting $\sqrt{\ \ }$ to $-\sqrt{\ \ }$ in the above.
Therefore the combined system $(x=x_1+x_2, p=p_1+p_2)$
may be viewed to be time-reversal invariant.

The Hamiltonian $H$ is from Eq. \g
\eqn\p{H=i\hbar(\eta_1N_1+\eta_2 N_2),}
which is indeed invariant under $U_{\alpha}(\alpha=t, \lambda)$ and
$N_i(i=1,2)$
are defined as follows
\eqn\aa{\eqalign{&N_i={1\over 2}(a_i b_i+b_i a_i)=-{1\over 2}(A_iB_i+B_iA_i)
\cr
		&a_i=iA_i, \ \ b_i=iB_i, \ \
            x_i=a_i\exp(\eta_i t), \ \ p_i={1\over 2}i\hbar
		b_i\exp(-\eta_i t).}}
One may easily verify the following commutation relations using the rules
\eqn\bb{\eqalign{&[a_i, a_j]=[b_i, b_j]=0, \ \ [a_i, b_j]=\delta_{ij} \cr
		&[A_i, A_j]=[B_i, B_j]=0, \ \ [B_i, A_j]=\delta_{ij} \cr
		&[N_i, a_j]=-a_i\delta_{ij}, \ \
		[N_i, b_j]=b_i\delta_{ij} \cr
		&[N_i, A_j]=-A_i\delta_{ij}, \ \
		[N_i, B_j]=B_i\delta_{ij} \cr
		&[N_1, N_2]=0.}}
{}From the relations \bb, we immediately have
\eqn\yyy{\eqalign{&N_i|n_i^{(\pm)}>=\pm(n_i^{(\pm)}+
		{1\over 2})|n_i^{(\pm)}> \cr
		&a_i|n_i^{(+)}>=\sqrt{n_i^{(+)}}|n_i^{(+)}-1>, \cr
		&b_i|n_i^{(+)}>=\sqrt{n_i^{(+)}+1}|n_i^{(+)}+1>, \cr
		&B_i|n_i^{(-)}>=\sqrt{n_i^{(-)}}|n_i^{(-)}-1>, \cr
		&A_i|n_i^{(-)}>=\sqrt{n_i^{(-)}+1}|n_i^{(-)}+1>, \cr
		&<n_i^{\alpha}|{n'}_i^{\alpha'}>=\delta_{n_i {n'}_i}\delta_{
		\alpha \alpha'} \ \ (\alpha=+, -) \cr
		&\ \ (n_i^{(\pm)}=0, 1, \cdots).}}
Therefore the Hamiltonian $H$ is exactly of
the harmonic oscillator type and has the same wave functions
(of course, with the completeness).

Note that since there is no prior reason to identify $a_i$ ($b_i$) as
the annihilation (creation) operators, we have used $A_i=ia_i$ and $B_i=ib_i$
(these transform the positions into the momenta and vice versa)
to obtain the second set of solutions. Namely $B_i$ ($A_i$) are
our new annihilation (creation) operators. Alternatively, we may simply
identify $A'_{i}=-a_{i}$ as new creation operators. All these
are similar to the usual case in which the canonical momentum operator
consists of the operations of both annihilation and creation.
Futhermore, it is worth pointing out that
we have a similar situation to Ref. [5]. An operator, typically
$A=xp+px, \ [x,p]=i\hbar$ (in our case, $x_ip_i+p_ix_i$), seems to
be hermitian but posses the pure imaginary eigenvalues. We resolve
this apparent contradiction in a similar way to that in Ref. [5], namely,
giving up the hermitian condition of $A$. Actually, an alternative
method may also be employed, in which one may directly solve
$(-2ixd_x-i)\hbar\psi=E\psi$ and assume $\psi$ to be single-value.
Immediately, one gets $E_n^{\pm}=\pm i(2n^{\pm}+1)\hbar$($n^{\pm}\ge 0$,
integer). The completeness
of the states will be followed as the coherent type weight functions
appear in the measure of the inner products.

The above discussion also agrees with the representation
theory as below. If one introduces
\eqn\kk{\eqalign{&J_1^i={1\over 4}(a_i^2+b_i^2) \cr
		&J_2^i={1\over 4}i(a_i^2-b_i^2) \cr
		&J_3^i={1\over 2}N_i,}}
one may easily verify that $J_{\beta}^i(\beta=1,2,3)$
are generators of the three dimensional
Lorentz group $SO(2, 1)$ or equivalently the two dimensional symplectic
group $Sp(2, R)$:
\eqn\ll{[J_1^i, J_2^i]=-iJ_3^i, \ \
[J_2^i, J_3^i]=iJ_1^i,\ \ [J_3^i, J_1^i]=iJ_2^i.}
According to Bargmann,$^{9}$ all irreducible unitary representations of
$SO(2,1)$
can be classified as follows: $D_\mu^{(+)}$ ($\mu<0$),
$D_\mu^{(-)}$ ($\mu<0$), principal
series, supplementary series, and the identity representation, where $\mu$
is the Casimir parameter, i.e. the Casimir operator $J_3^2-J_1^2-J_2^2=
\mu(\mu+1)$. In terms of the $A's$ and $B's$, there will
an additional minus sign in the $J's$. Because of the sign change
of the $J_3^i$, $-J_3^i={1\over 2}(A_iB_i+B_iA_i)$ will have the
bounded from below spectrum as the square integrable wave functions
are chosen. Thus,
$N^{(\pm)}=\pm(n+{1\over 2})$ as $\mu=\mp {1\over 4}$ ($\mu$ is determined
by the commutation relations of $a's$ and $b's$ or $A's$ and
$B's$.), respectively.
Note that for other values of $\mu$, one obtains the representations
for one dimensional anyons,$^{10}$ or for parabosons,$^{11}$ depending on
one's viewpoint.

Clearly, the spectrum of $H$ contains all information of
the damped system $x_i$ and its time reversed partner(growing one)
$p_i$, as one expects. However, the spectrum of $H$ is lost the usual
meaning of energy levels and $H$ may only be viewed as the dynamical
generator. To be more specific, we focus on the case of the underdamping,
$\lambda<\omega$. In this case, the spectrum of $H$ can be explicitly
displayed as
\eqn\op{\eqalign{&E_{n_1 n_2}^{(+)}=\hbar\Omega(n_2^{(+)}-
		n_1^{(+)})-i\hbar\lambda(n_1^{(+)}+n_2^{(+)}+1) \cr
		&E_{n_1 n_2}^{(-)}=\hbar\Omega(n_1^{(-)}-
		n_2^{(-)})+i\hbar\lambda(n_1^{(-)}+n_2^{(-)}+1),}}
where $\Omega=\sqrt{\omega^2-\lambda^2}$ and $E_{n_1 n_2}^{(+)}$ describe
decaying states while $E_{n_1 n_2}^{(-)}$ describe growing states.
One may easily demonstrate that the states with $E^{(+)}$ is indeed
the time reverse of the states with $E^{(-)}$. Our results \op \
agree with those appeared in Ref. [5].
Since the well-known harmonic oscillator wave functions are
complete, the time-evolution of the Schr${\ddot{\rm o}}$dinger
equation can be solved without trouble and of course is no longer
unitary due to the damping force. For the harmonic oscillator,
we see that the states of $H$ becomes those of the usual undamped
harmonic oscillator in the limit $\lambda\rightarrow 0$ if only
those states, satisfying the conditions $a_1|\psi>=b_2|\psi>=0$
are considered. This is similar to neglect the negative
energy states which are physically unstable as we interpret
$H$ as the energy operator.

Finally, we explicitly present here the Schr$\ddot{\rm o}$dinger
time evolution under $H$ for a given initial state
$|\psi(0)>=\sum_{n_i^{(\pm)}} c_{n_1^{(\pm)}n_2^{(\pm)}}|n_1^{(\pm)}
n_2^{(\pm)}>$,
\eqn\xx{\eqalign{|\psi(t)>&=\sum_{n_i^{(\pm)}} c_{n_1^{(\pm)}n_2^{(\pm)}}
		e^{-{i\over\hbar}H t}|n_1^{(\pm)}n_2^{(\pm)}>  \cr
		&=\sum_{n_i^{(\pm)}} c_{n_1^{(\pm)}n_2^{(\pm)}}
		e^{-{i\over\hbar}E_{n_1 n_2}^{(\pm)} t}
		|n_1^{(\pm)}n_2^{(\pm)}>,}}
where $E_{n_1 n_2}^{(\pm)}$ are spectrum of $H$.
Alternatively, one may easily write down the Schr$\ddot{\rm o}$dinger
time evolution of an initial state $|phi(0)>=
\sum_{n_i^{(+)}} c_{n_1 n_2}^{(+)}|n_1^{(+)}
n_2^{(+)}>$ which decays with time
\eqn\xxxx{|\phi(t)>=\sum_{n_i^{(+)}}c_{n_1 n_2}^{(+)}
e^{-i\Omega(n_2^{(+)}-
n_1^{(+)})-\lambda(n_1^{(+)}+n_2^{(+)}+1)}|n_1^{(+)} n_2^{(+)}>.}

The time dependent operator expectation values can
be obtained from either the equations
of motion or from the standard Heisenberg representations. For example,
\eqn\zz{\eqalign{\bar{x}_i&=<\psi(t)|x_i|\psi(t)> \cr
	&=<\psi(0)|x_i(t)|\psi(0)>=
<\psi(0)|a_i|\psi(0)>e^{\eta_i t},}}
where by definition $x_i=\exp(iH t/\hbar)a_i\exp(-iH t/\hbar)$
(from the Heisenberg equation \j).
Note the definition of the ``bra" $<\psi(t)|$ is changed to
the time reversed conjugation of $|\psi(t)>$, rather than
the usual hermitian conjugation of $|\psi(t)>$. This is natural
since the Hilbert space is defined by our time reversal invariant
operator $H$. The same definition of the inner products
appeared also in Ref. [5].
Likewise, one may easily write down
the correlations
\eqn\zzz{<\psi(0)|x_1(t)x_2(t')|\psi(0)>=
<\psi(0)|a_1a_2|\psi(0)>e^{\eta_1 t +\eta_2 t'}.}

In conclusion, we have analyzed the linear damped harmonic oscillator
using the Hamiltonian framework, which is naturally introduced for
describing this system and its time reversed partner.
We have shown that the conventional
Schr$\ddot{\rm o}$dinger dynamics is explicitly realized.
Furthermore, the Schr$\ddot{\rm o}$dinger wave description has the usual
interpretation and
our method may be easily generalized to quantize three dimensional
damped systems. Second quantization may also be carried out
so that a field theory for this model is realized.

Since our analysis starts directly by quantizing the classical solutions
or phase space, as we have demonstrated, our method may yield a
straightforward framework to study
general non-Hamiltonian systems. We outline
here how this may work. One will need first to identify the asymmetries
which are responsible for the non-existence of the Hamiltonian of a
system. These asymmetries are usually related to the time-reversal
transformations because non-Hamiltonian systems
have the common feature of history dependence. Then
to find the classical solutions of both the systems will be
the next step.
Finally, one may find the Hamiltonian in which the original
asymmetries will be restored to describe the combined system
and the canonical quantization will be carried out directly.
One such immediate example is the field theory of anyons.
In 2+1 dimensions, the positive and negative spin corresponds to
the different models$^{1,2}$ and they are related by the time reversed
symmetry (similar to the magnetic
like interactions),$^{2}$ one may hope that our method may be used
for quantizing anyonic field theories in a similar manner.

\vskip 0.2in
\noindent
{\bf Acknowledgements}
\vskip .1in
We thank V. P. Nair and H. C. Ren for fruitful discussions, and
M. D. Doyle and V. P. Nair for reading the manuscript.

\vskip 0.2in
\noindent
{\bf References}
\vskip .1in
\item{1.} {R. Jackiw and V. P. Nair, {\it Phys. Rev.} {\bf D43}, 1993 (1991);
	M. S. Plyushchay, {\it Phys. Lett.} {\bf B248}, 107 (1990); D. Shon
	and S. Khlebnikov, {\it JETP Lett.} {\bf 51}, 611 (1990).}
\vskip .1in
\item{2.} {C. Chou, V. P. Nair, and A. P. Polychronakos,
	RU-92-17-B/CU-TP-584/CERN-TH 6768/93(January, 1993)
	(to appear in {\it Phys. Lett.} {\bf B}).}
\vskip .1in
\item{3.} {V. Bargmann and E. P. Wigner, {\it Proc. Nat. Acad. Sci. U.S.A.}
	{\bf 34}, 211 (1948).}
\vskip .1in
\item{4.} {H. Dekker, {\it Z. Phys.} {\bf B21}, 295 (1975); {\it Phys. Rev.}
	{\bf A16}, 2126 (1977).}
\vskip .1in
\item{5.} {H. Feshbach and Y. Tikochinsky, {\it A Festschrift for I. I.
	Rabi} ({\it Transactions New York Academy
	of Sciences}, Series II, Vol. 38), edited by L. Motz,  p44-53
	(New York, NY 1977).}
\vskip .1in
\item{6.} {For example, I. R. Senitzky, {\it Phys. Rev.} {\bf 119}, 670
(1960).}
\vskip .1in
\item{7.} {M. D. Kostin, {\it J. Chem. Phys.} {\bf 57}, 3589 (1972);
	A. P. Polychronakos and R. Tzani,
	CU-TP-569/UB-ECM-PF-92/16(June, 1992).}
\vskip .1in
\item{8.} {C. Chou, in preparation.}
\vskip .1in
\item{9.} {V. Bargmann, {\it Ann. Math.} {\bf 48}, 568 (1947).}
\vskip .1in
\item{10.} {J. M. Leinaas and J. Myrheim, {\it Nuo. Cim.} {\bf 37B}, 1 (1977);
		{\it Int. J. Mod. Phys.} {\bf B5}, 2573 (1991);
		ISSN 0365-2459 (November, 1992).}
\vskip .1in
\item{11.} {For example, Y. Ohnuki and S. Kamefuchi, {\it Quantum Field
	Theory and Parastatistics} (Tokyo, 1982).}
\end